\documentclass[useAMS,usenatbib]{mn2e}
\usepackage{graphicx}
\usepackage{epsfig}
\usepackage{epstopdf}
\usepackage{amssymb}

\title[Model of the vibrationally excited H$_2$O maser at 658 GHz] {Model of the vibrationally excited H$_2$O maser at 658~GHz in circumstellar envelopes around asymptotic giant branch stars}

\author[A.V. Nesterenok] {A.V. Nesterenok $^{1}$ \thanks{E-mail:alex-n10@yandex.ru} \\
  $^{1}$ Ioffe Institute, 26 Polytechnicheskaya St., 194021 Saint Petersburg, Russia}

\begin{document}

%\date{Accepted . Received ; in original form }
\pagerange{\pageref{firstpage}--\pageref{lastpage}} %\pubyear{2015}
\maketitle

\label{firstpage}

\begin{abstract}
The model is presented of H$_2$O maser in the $1_{10}-1_{01}$ line within the first excited vibrational state of the molecule around oxygen-rich asymptotic giant branch stars. It is suggested that the maser cloud is located in the inner layers of the circumstellar envelope where intense dust formation takes place. The calculations took into account rotational levels belonging to the five lowest vibrational states of the H$_2$O molecule. The model predicts the gain values of the 658-GHz maser about $10^{-14} -10^{-13}$~cm$^{-1}$ at H$_2$ molecule concentrations $10^9 - 10^{11}$~cm$^{-3}$ and at high ortho-H$_2$O concentrations $ \ga 10^5$~cm$^{-3}$. The gas temperatures 1000 -- 1500~K are considered to be a necessary condition for the effective maser operation. Results are presented for other maser transitions of the excited vibrational states of the molecule.
\end{abstract}


\section{Introduction}

Intermediate- and low-mass stars (with main sequence mass 1 -- 6~M$_\odot$) cease their life on the asymptotic giant branch (AGB), where they combine a high luminosity (i.e. 3000~L$_{\odot}$ and higher) and a low effective temperature, below 3000~K \citep{Habing1996}. Stars with higher mass become red supergiants and have ten times the size of AGB stars. Late-type stars lose a substantial fraction of their initial mass by stellar winds and are surrounded by extended envelopes. Masers provide an unique probe of structure and kinematics of the circumstellar environment of late-type stars. Maser emission from the three most common maser molecules -- SiO, H$_2$O, and OH -- traces regions of the circumstellar envelopes of oxygen-rich evolved stars on scales from a few to several hundred astronomical units, hereafter AU \citep{Alcolea2004}. 

Intense H$_2$O maser emission in the 22.2-GHz $6_{16} \to 5_{23}$ transition\footnote{Energy levels of the asymmetric rotor molecule H$_2$O are denoted $J_\rmn{K_a K_c}$ where $J$ is the total rotational quantum number, $K_\rmn{a}$ and $K_\rmn{c}$ represent its projections along the molecular axes. The vibrational states are denoted by quantum numbers $(\nu_1, \nu_2, \nu_3)$, where $\nu_1$ denotes symmetric stretch, $\nu_2$ -- bending, and $\nu_3$ -- antisymmetric stretch, respectively.} belonging to the ground vibrational state of the molecule has been found towards many hundreds of circumstellar envelopes of Mira and semi-regular long-period variable stars and red supergiants \citep{Valdettaro2001, Richards2012}. Radio interferometry methods allow us to image H$_2$O masers in this line at high resolution and the structure of the circumstellar envelopes can be studied in detail. In addition to the 22.2-GHz line, maser emission from a number of other ortho- and para-H$_2$O (sub)millimetre wavelength transitions belonging to the ground and first excited vibrational states has been detected \citep{Humphreys2007, Hunter2007}. The maser emission in rotational lines within the excited vibrational states is of particular interest. Given the excitation conditions, these lines must arise from the innermost hot regions of circumstellar envelopes.

The emission from two rotational lines within the first excited vibrational state (010) of H$_2$O was discovered by \citet{Menten1989} using the Institut de Radioastronomie Millim$\acute{\mathrm{e}}$trique (IRAM) telescope. The para-H$_2$O transition $4_{40} \to 5_{33}$ at a frequency of 96~GHz and ortho-H$_2$O transition $5_{50} \to 6_{43}$ at 233~GHz were detected towards the supergiant VY CMa, the latter transition was also detected towards the semiregular variable W Hya. The emission at 96~GHz towards VY CMa showed signs of maser action. 

\citet{Menten1995} detected strong H$_2$O maser emission at a frequency of 658~GHz towards a sample of 10 oxygen-rich evolved stars using the 10.4-m telescope of the Caltech Submillimeter Observatory. The line corresponds to the $1_{10}-1_{01}$ rotational transition within the (010) excited vibrational state of the ortho-H$_2$O molecule. The 658-GHz spectra show contiguous, broad, emission with asymmetric line shapes centred at the stellar velocity. The photon luminosities of the 658-GHz masers, assuming the isotropic emission, ranged from $5 \times 10^{43}$ to $3 \times 10^{46}$~s$^{-1}$. For the detected sources the photon luminosity of the 658-GHz line exceeded that of the 22.2-GHz line and the SiO maser lines. \cite{Hunter2007} reported the  observations of the 658-GHz circumstellar masers obtained by the Submillimeter Array. They presented new detections from six Mira variables including IK Tau. \citet{Richards2014} presented Atacama Large Millimeter/submillimeter Array (ALMA) observations of the 658-GHz water maser towards VY CMa. The interferometric observations with high angular resolution enabled us to locate water masers with milliarcsecond precision. The 658-GHz masers have a complex distribution close to the star and are located mostly outside the outer rim of SiO masers in this source. The 658-GHz maser emission is more concentrated towards the stellar position than the emission in the other H$_2$O maser lines at 22.2~GHz, 321~GHz and 325~GHz \citep{Richards2014}.

Other lines of the (010) excited vibrational state of the H$_2$O molecule were sought. \citet{Feldman1993} detected the $5_{23} \to 6_{16}$ rotational line of ortho-H$_2$O at 336~GHz towards VY CMa using the James Clerk Maxwell Telescope. \citet{Menten2006} reported the detection of the emission in the ortho-H$_2$O lines $6_{61} - 7_{52}$ at 294~GHz and $5_{23} \to 6_{16}$ at 336~GHz towards VY CMa using the Atacama Pathfinder Experiment (APEX) telescope. Searches have been performed for three transitions with negative results: para-H$_2$O transition $6_{60} \to 7_{53}$ at 297~GHz \citep{Menten2006}, ortho-H$_2$O $4_{23} \to 3_{30}$ at 12.0~GHz \citep{Myers1982}, and ortho-H$_2$O $4_{14} \to 3_{21}$ at 67.8~GHz \citep{Petuchowski1991}.

\citet{Deguchi1977} included the vibrationally excited state (010) in calculations of the pumping of H$_2$O masers in circumstellar envelopes of late-type stars. His model predicted maser emission in 12.0-GHz, 67.8-GHz and other lines belonging to this excited vibrational state. \citet{Alcolea1993} predicted maser action in several H$_2$O transitions belonging to the first excited vibrational state whose upper level quantum numbers are $K_\rmn{a} = J$ and $K_\rmn{c} = $ 0 or 1. However, the maser action in the 658-GHz line was not predicted in the calculations of \citet{Deguchi1977} and \citet{Alcolea1993}. This maser line was also not mentioned in the study of circumstellar H$_2$O excitation by \citet{Deguchi1990}.

Until recently the modelling of the vibrational excitation of H$_2$O molecule was hampered by the lack of collisional excitation rates for transitions between the ground vibrational state and excited vibrational states. \citet{Faure2008} presented collisional rate coefficients for the (de)excitation by H$_2$ of the lowest 824 ro-vibrational levels of the H$_2$O molecule in the temperature range $200-5000$~K. These data can be used for modelling the water molecule excitation in warm astrophysical media.

In this paper we study the physical conditions necessary for the generation of strong maser emission in the vibrationally excited H$_2$O line at 658~GHz in the circumstellar envelopes around oxygen-rich AGB stars. In our model the 658-GHz maser is located in the dust formation zone in the inner layers of the circumstellar envelope.

\section{Physical parameters of the model}
\subsection{Inner region of circumstellar envelopes}
The star's physical parameters adopted in our calculations correspond to those of the AGB star IK Tau (Table \ref{table1}). IK Tau is an M-type star and is at a distance of about 250~pc \citep{Olofsson1998}. The angular diameter of the stellar photosphere was estimated by \citet{Monnier2004} to be 20~mas based on the combined Keck aperture-masking data and Infrared Optical Telescope Array (IOTA) data.

IK Tau exhibits circumstellar maser emission in SiO, H$_2$O, and OH molecules. The SiO maser emission is found to arise from a clumpy ring within two to three stellar radii from the star centre \citep{Boboltz2005, Matsumoto2008}. The emission of the vibrationally excited H$_2$O masers and SiO masers arises from a similar range of velocities, which can be interpreted as evidence for a close physical origin \citep{Menten1991, Menten1995, Hunter2007}. The observations of the variability of the SiO maser distribution reveal that the medium in the inner layers of circumstellar envelopes is strongly turbulent. The turbulent velocities of the gas can be as high as several km~s$^{-1}$ \citep{Tsuji1986, Decin2012}. 

Dynamical models of the AGB star winds predict shocks and large velocity gradients up to several km~s$^{-1}$~AU$^{-1}$ in the inner layers of circumstellar envelopes \citep{Woitke2006, Nowotny2010, Ireland2011}. The gas temperature in these layers tends to be close to the radiative equilibrium temperature, except for some narrow, hot post-shock cooling zones \citep{Schirrmacher2003}.

Hydrogen is contained mainly in the form of H$_2$ molecules in the photosphere of cool late-type stars \citep{Glassgold1983}. Hydrogen molecules are destroyed by the dissociative shock waves propagating through the stellar atmosphere and are then re-formed on dust grains. For the sake of simplicity, the hydrogen is considered to be molecular in the calculations. The relative abundance of H$_2$O molecules in the stellar wind depends on the carbon-to-oxygen ratio in the stellar photosphere and can vary over a wide range \citep{Cherchneff2006}. 

\begin{table}
\caption{Physical parameters of the model}
\begin{tabular}{ll}
\hline \\ [-2ex]
Stellar radius  & $R_*$ = 2.5~AU \\ [5pt]
Stellar temperature$^{1}$ & $T_*$ = 2300~K \\ [5pt]
Distance from star centre$^{2}$ & $D$ = 12.5~AU \\ [5pt]
Cloud thickness$^{2}$ & $H$ = 1.25~AU \\ [5pt]
Number density of H$_2$ & $3 \times 10^8 \leq N_{\rmn{H_2}} \leq 10^{11}$~cm$^{-3}$ \\ [5pt]
Number density of ortho-H$_2$O & $10^4 \leq N_{\rmn{H_2O}} \leq 2 \times 10^6 $~cm$^{-3}$ \\ [5pt]
Gas temperature & 600 $\leq T_\rmn{g} \leq$ 1600~K \\ [5pt]
Microturbulent velocity & $v_{\rmn{turb}}$ = 3~km~s$^{-1}$ \\ [5pt]
Velocity gradient & $k_\rmn{v}$ = 5~km~s$^{-1}$~AU$^{-1}$ \\ [5pt]
Dust--gas mass ratio & $0 \leq f_\rmn{d} \leq 0.005$ \\ [5pt]
\hline \\ 
\end{tabular}
$^{1}$ The effective temperature of the photosphere of IK Tau \citep{Monnier2004}. $^{2}$ See Fig. 1.
\label{table1}
\end{table}

\subsection{Dust model}
The mass loss observed in evolved AGB stars is usually attributed to a two-stage process: atmospheric levitation by pulsation-induced shock waves followed by radiative acceleration of dust grains, which transfer momentum to the surrounding gas through collisions \citep{Bladh2012}. In order for a dust species to trigger a wind it has to form close to the stellar surface, within reach of the shock waves. Radiation-hydrodynamical models of winds of oxygen-rich M-type AGB stars suggest that the winds can be driven by photon scattering on Fe-free silicate grains of sizes 0.1-1~$\umu$m \citep{Hofner2008, Bladh2012}. The observations of three M-type AGB stars using the multi-wavelength aperture-masking polarimetric interferometry pointed to the presence of dust shells with large grains (about 0.3~$\umu$m in radius) at radii of about two stellar radii \citep{Norris2012}. The inner boundary of the dust formation zone appears to be located in or just outside the SiO maser zone.

In our model, the forsterite Mg$_2$SiO$_4$ was considered as the dust grain material \citep{Bladh2012}; some calculations were done for magnesium-iron silicates. Here we used the refractive index data for dust from \citep{Jager2003, Dorschner1995} \footnote{http://www.astro.uni-jena.de/Laboratory/Database/databases.html}. In our calculations, the dust particle radius $a$ was assumed to be 0.3~$\umu$m. The cross sections for the absorption of radiation by dust particles were calculated using the Mie scattering theory. We used the numerical code published in the monograph by \citet{Bohren1983} and modified by \citet{Draine2004} \footnote{http://code.google.com/p/scatterlib/wiki/Spheres}. The dust grains grow until the elements contributing to the dust material are consumed due to the relative element abundances in the stellar atmosphere. The first element in Mg$_2$SiO$_4$ that will be completely consumed is Mg for a solar abundance of elements. The value of the dust--gas mass ratio $f_\rmn{d}$ determined by the relative abundance of Mg is approximately 0.002. The parameter $f_\rmn{d}$ was varied over the range 0 -- 0.005 in the calculations.

The dust heating mechanism in the inner layers of the circumstellar envelopes is the absorption of stellar radiation. The dust loses thermal energy through emission of radiation. The dust emissivity $\varepsilon_\rmn{c}(\nu)$ was calculated in accordance with Kirchhoff's radiation law; the function $\varepsilon_\rmn{c}(\nu)$ strongly depends on dust temperature $T_\rmn{d}$. The dust temperature $T_\rmn{d}$ at distance $D$ from the stellar centre was calculated from the solution of the equation of thermal balance:

\begin{equation}
\displaystyle
4\pi \int \limits_0^{\infty} d \nu \, \varepsilon_\rmn{c} (\nu) = \Omega_* \int \limits_0^{\infty} d \nu \, \kappa_\rmn{c}(\nu) I_*(\nu),
\label{dust_temp_eq}
\end{equation}

\noindent
where $\kappa_\rmn{c}(\nu)$ is the dust absorption coefficient, $I_*(\nu)$ is the intensity of stellar radiation, $\Omega_*$ is the solid angle that cuts out the stellar disk on the celestial sphere at distance $D$. Here, an optically thin stellar atmosphere is assumed, where the incident intensity on the grains is direct star light.

\begin{figure}
\includegraphics[width=84mm]{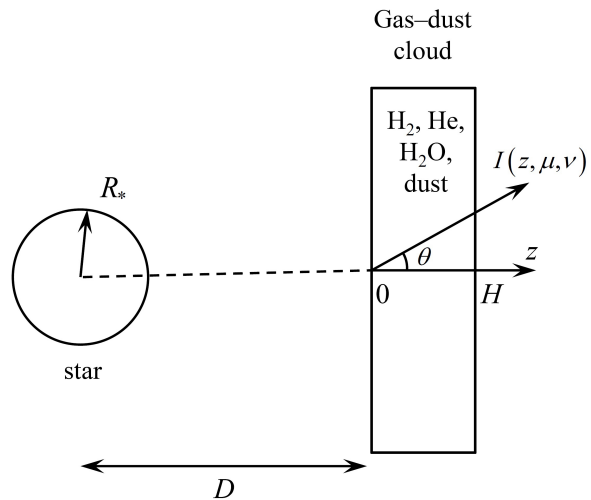}
\caption{Gas--dust cloud model. The distance from the maser cloud to the star centre is equal to $D = 5R_*$. }
\label{fig1}
\end{figure}

\subsection{Maser cloud}
There is no interferometric data for the (sub)millimetre H$_2$O masers with such resolution that individual maser clouds can be resolved \citep{Richards2014}. We suggest that vibrationally excited H$_2$O maser radiation originates in compact gas clouds in the warm inner layers of circumstellar envelopes. The maser cloud dimensions were set to be equal to 0.1 of the distance from the parent star. Such cloud sizes were revealed for the 22.2-GHz H$_2$O masers based on radio interferometry observations \citep{Richards2012}. 

We consider the one-dimensional model of a flat gas--dust cloud (see Fig. \ref{fig1}). The $z$ coordinate axis passes through the star centre and is perpendicular to the cloud plane. The cloud sizes along two coordinate axes are much larger than those along the third $z$ coordinate axis. We assume that there is a gas velocity gradient along the $z$ axis, $k_\rmn{v} = dv/dz > 0$. The cloud consists of a mixture of H$_2$ and H$_2$O molecules, He atoms, and dust particles and is in the radiation field of the parent star. For simplification, the physical parameters of the cloud (the gas and dust temperatures, the number densities of atoms and molecules, and the dust abundance) were assumed to be independent of the coordinates. However, the H$_2$O level populations were considered as functions of the $z$ coordinate. We assume that $H << D$ (plane geometry).

The gas temperature in the maser clouds must satisfy $T_\rmn{g} \la 1600$~K, because there is dissociation of H$_2$O molecules at higher temperatures \citep{Jeong2003, Nesterenok2011}. The model parameters adopted in our calculations are given in the Table \ref{table1}.

\section{Calculation of the H$_2$O level populations}
\subsection{System of master equations for the molecule level populations}
In the stationary case, the system of equations for the level populations is

{\setlength{\mathindent}{0pt}
\begin{equation}
\begin{array}{l}
\displaystyle
\sum_{k=1, \, k \ne i}^M \left( R_{ki}(z)+C_{ki} \right) n_k(z) - \\ [15pt]
\displaystyle
- n_i(z) \sum_{k=1, \, k \ne i}^M \left( R_{ik}(z)+C_{ik} \right)=0, \quad i=1,...,M-1, \\ [15pt]
\displaystyle
\sum_{i=1}^M n_i(z)=1,
\end{array}
\label{stat_eqns}
\end{equation}}

\noindent
where $n_i$ is the normalized population of level $i$, $M$ is the total number of levels, $R_{ik}(z)$ is the rate coefficient for the transition from level $i$ to level $k$ through radiative processes, and $C_{ik}$ is the rate coefficient for the transition from level $i$ to level $k$ through collisional processes. The rate coefficients for radiative transitions $R_{ik}(z)$ are

\begin{equation}
\begin{array}{l}
R_{ik}^{\downarrow}(z)=B_{ik}J_{ik}(z)+A_{ik}, \quad \varepsilon_i > \varepsilon_k, \\[10pt]
R_{ik}^{\uparrow}(z)=B_{ik}J_{ik}(z), \quad \varepsilon_i < \varepsilon_k,
\end{array}
\nonumber
\end{equation}

\noindent
where $A_{ik}$ and $B_{ik}$ are the Einstein coefficients for spontaneous and stimulated emission, respectively; $\varepsilon_i$ and $\varepsilon_k$ are the energies of the levels $i$ and $k$, respectively; $J_{ik}(z)$ is the radiation intensity averaged over the direction and over the line profile. For the plane parallel geometry:

\begin{equation}
J_{ik}(z)=\frac{1}{2} \int\limits_{-\infty}^{\infty} d\nu \int\limits_{-1}^{1} d\mu \; \phi_{ik}(z,\mu,\nu) I(z,\mu,\nu).
\nonumber
\end{equation}

\noindent
Here, $I(z,\mu,\nu)$ is the intensity of radiation at a frequency $\nu$ in direction $\mu = cos \theta$, where $\theta$ is the angle between the $z$ axis and the radiation direction, $\phi_{ik}(z,\mu,\nu)$ is the normalized spectral line profile. The line intensity $I(z,\mu,\nu)$ is the solution of equation of radiative transfer:

\begin{equation}
\mu \frac{dI(z,\mu,\nu)}{dz} = -\kappa(z,\mu,\nu) I(z,\mu,\nu) + \varepsilon(z,\mu,\nu),
\label{rad_tr}
\end{equation}

\noindent
where $\varepsilon(z,\mu,\nu)$ is the emission coefficient, and $\kappa(z,\mu,\nu)$ is the absorption coefficient. Each of the coefficients $\varepsilon(z,\mu,\nu)$ and $\kappa(z,\mu,\nu)$ is the sum of the emission or absorption coefficient of the dust and the emission or absorption coefficient in a spectral line, respectively:

\begin{equation}
\begin{array}{l}
\displaystyle
\varepsilon(z,\mu,\nu) = \varepsilon_\rmn{c}(\nu)+\frac{\displaystyle h\nu}{4\pi}A_{ik} N n_i(z)\phi_{ik}(z,\mu,\nu), \\ [10pt]
\displaystyle
\kappa(z,\mu,\nu) = \kappa_\rmn{c}(\nu) + \\ [10pt]
\displaystyle
+ \frac{\lambda^2}{8\pi}A_{ik}N \left (\frac{g_i}{g_k}n_k(z)-n_i(z) \right) \phi_{ik}(z,\mu,\nu).
\end{array}
\end{equation}

\noindent
Here, $N$ is the molecule number density; $g_i$ and $g_k$ are the statistical weights of the levels; $\lambda$ is the radiation wavelength. In these formulas, it is implied that level $i$ lies above level $k$ in energy, i.e., $\varepsilon_i > \varepsilon_k$. The spectral profile of the emission and absorption coefficients in the laboratory frame of reference is:

\begin{equation}
\displaystyle
\phi_{ik}(z, \mu, \nu) = \tilde{\phi}_{ik}(\nu - \mu \nu_{ik} v(z)/c)
\nonumber
\end{equation}

\noindent
where $\tilde{\phi}_{ik}(\nu)$ is the normalized spectral line profile in the co-moving frame of the gas, $\nu_{ik}$ is the transition frequency, $v(z)$ is the gas velocity along the $z$ axis, here $v(z) = k_\rmn{v}z$, $k_\rmn{v}$ is a constant coefficient equal to the gas velocity gradient in the cloud. 

The normalized spectral line profile in the co-moving frame of the gas is:

\begin{equation}
\tilde{\phi}_{ik}(\nu) = \frac{1}{\sqrt{\pi}\Delta \nu_{ik}} e^{-x^2}, \quad x = (\nu - \nu_{ik})/\Delta \nu_{ik}
\label{x_def}
\end{equation}

\noindent
where $\Delta \nu_{ik}$ is the line profile width. The line profile width is determined by the spread in thermal velocities of molecules and turbulent velocities in the gas--dust cloud:

\begin{equation}
\Delta\nu_{ik}=\nu_{ik}\frac{v_{\rmn{D}}}{c}, \quad v_{\rmn{D}}^{2} = v_{\rmn{T}}^2 + v_{\rmn{turb}}^2,
\nonumber
\end{equation}

\noindent
where $v_T = \sqrt{2kT_{\rmn{g}}/m}$ is the most probable thermal velocity of the molecules, $k$ is the Boltzmann constant, $m$ is the mass of the molecule, and $v_{\rmn{turb}}$ is the characteristic turbulent velocity in the cloud.

Let $\mu_\rmn{c}$ to be the cosine of the angle between the $z$ axis and the radiation direction from the edge of the stellar disk, $\mu_\rmn{c} = \sqrt{1 - R_*^2/D^2}$. The boundary condition for eq. (\ref{rad_tr}) at $z = 0$ is $I(0,\mu,\nu) = I_*(\nu)$, $\mu \geq \mu_\rmn{c}$, and $I(0,\mu,\nu) = 0$, $\mu_\rmn{c} > \mu \geq 0$. The star is assumed to have a black-body radiation spectrum with photospheric temperature equal to 2300~K. At $z = H$ we take the boundary condition to be $I(H,\mu,\nu) = 0$, $\mu < 0$. 

There is no line overlap between the transitions of the H$_2$O molecule. The emission from inverted transitions is disregarded in our calculations of the energy level populations of the molecule.

\subsection{Spectroscopic data and collisional rate coefficients}
In our calculations, we took into account 411 rotational levels of ortho-H$_2$O and 411 rotational levels of para-H$_2$O belonging to the five lowest vibrational states of the ground electronic state of the molecule. The first rotational levels of the excited vibrational states have the following energies: 1594.8~cm$^{-1}$ for the (010) vibrational state, 3151.6~cm$^{-1}$ for (020), 3657.1~cm$^{-1}$ for (100), and 3755.9~cm$^{-1}$ for (001). The spectroscopic data for H$_2$O molecule were taken from the HITRAN 2012 database \citep{Rothman2013}. The energies of the rotational-vibrational transitions for the H$_2$O levels under consideration are less than 4400~cm$^{-1}$ (2.3~$\umu$m) and correspond to the infrared and the radio band. Note that the peak of the intensity of black-body radiation $B(\nu)$ at a temperature of $T_* = 2300$~K is near 2.2~$\umu$m.

The rate coefficients for collisional transitions between H$_2$O levels in collisions of H$_2$O with H$_2$ were taken from \citet{Faure2007} and \citet{Faure2008}. The rate coefficients for collisional transitions between H$_2$O levels in inelastic collisions of H$_2$O with He atoms were taken from \citet{Green1993}. The data from \citet{Green1993} contain the collisional rate coefficients for the lowest 45 levels of ortho-H$_2$O molecule and 45 levels of para-H$_2$O. The collisional rate coefficients for the transitions including higher H$_2$O levels were calculated by extrapolating the data from \citet{Green1993} using the algorithm proposed by \citet{Faure2008} for the rate coefficients for H$_2$O and H$_2$ collisions. In our calculations, we used the relaxation rates of excited vibrational states of H$_2$O by He from \citet{Kung1975}. The ratio of the summed collisional rate coefficients out of any H$_2$O level for H$_2$O -- He and H$_2$O -- H$_2$ collisions lies in the range 0.1 -- 1. The abundance of helium atoms relative to the total number density of hydrogen nuclei was assumed to be $N_\rmn{He}/2N_\rmn{H_2} = 0.1$. Thus, the contribution of H$_2$O -- He collisions to the collisional excitation of H$_2$O molecule is about 10 per cent.

\subsection{Numerical calculations}
The system of master equations for the H$_2$O level populations eq. (\ref{stat_eqns}) and radiative transfer equation in the medium, eq. (\ref{rad_tr}), form a system of non-linear equations for the level populations of the molecule. This system of equations was solved by the accelerated $\Lambda$-iteration method \citep{Rybicki1991}. The Feautrier method was used for solving the radiative transfer equation \citep{Feautrier1964, Peraiah2004}. The collisional and radiative transitions between ortho and para spin isomers of H$_2$O molecule are forbidden, and the level populations of the molecules are calculated independently.

The cloud in our numerical model was broken down into layers parallel to the cloud plane. The molecular level populations within each layer were constant. The thickness of the near-surface layer was chosen to be $\Delta z_s = 10^{11} N_\rmn{H_2O}^{-1}$~cm. The optical depth for any H$_2$O line and for any considered direction in the near-surface layer was less than 1. The thickness of each succeeding layer into the cloud was larger than that of the preceding one by a constant factor. The number of layers into which the cloud was broken down is 100. The calculations showed that the decrease of the surface layer thickness (by a factor of 10) or the increase of the number of layers lead to a small effect on the population inversion of the transitions in question ($\la 1$ per cent).

The intensity is averaged over opposite directions in the Feautrier calculational scheme; the range of values for the parameter $\mu$ is $[0;1]$. The number of angles used in angular integration was set to be 15. The minimal value of the parameter $\mu$ for which the radiative transfer equation was solved is equal to 0.05, the corresponding $d\mu$ is equal to 0.1. The maximal value of the parameter $\mu$ is equal to 1 and corresponding $d\mu = 1-\mu_\rmn{c}$. Other angle values are distributed uniformly over the interval. The range of values for the parameter $x$ for each line (see eq. \ref{x_def}) was chosen to be $[-5;10]$; the centre of the interval is shifted to positive values in order to take into account the large velocity gradients in the cloud ($k_\rmn{v} \geq 0$). The discretization step of the parameter $x$ is 0.2. The model was tested with the number of frequency or angle points twice as much as listed above; the changes in the population inversions were negligible ($\sim 0.1$ per cent).

An additional acceleration of the iterative series was achieved by applying the convergence optimization method proposed by \citet{Ng1974}. The convergence criterion for the iterative series was the condition on the maximum relative increment in level populations for two successive iterations, 

\begin{equation}
\displaystyle \max_i |\Delta n_i/n_i| < 10^{-4}.
\end{equation}

\noindent
The error of the population inversion corresponding to this convergence criterion is $\la 1$ per cent for the parameter values of physical interest.

In our calculations, we used the algorithms for solving systems of linear equations published in \citet{Rybicki1991} and in the book by \citet{Press1997}. The calculations were performed on the supercomputer of the Saint Petersburg branch of the Joint Supercomputer Center of the Russian Academy of Sciences\footnote{http://scc.ioffe.ru/}.

\section{Results}
\subsection{Gain of the 658-GHz maser line}
The expression for the gain of the transition $i \to j$ at the line centre in a direction along the cloud plane is

\begin{equation}
\displaystyle
\gamma_{ij}(z)=\frac{\lambda^2 A_{ik} N}{8 \pi \sqrt{\pi} \Delta \nu_{ij}} \left(n_i(z)-\frac{g_i}{g_k}n_k(z) \right).
\nonumber
\end{equation}

\noindent
The cloud-averaged gain is calculated from the formula 

\begin{equation}
\overline{\gamma} = \frac{1}{H} \int\limits_{0}^{H} dz \, \gamma(z).
\end{equation}

\noindent
Inverted transitions with a small gain $\overline{\gamma} < 10^{-2} H^{-1}$~cm$^{-1}$ were ignored in the result analysis.

\begin{figure}
\includegraphics[width=84mm]{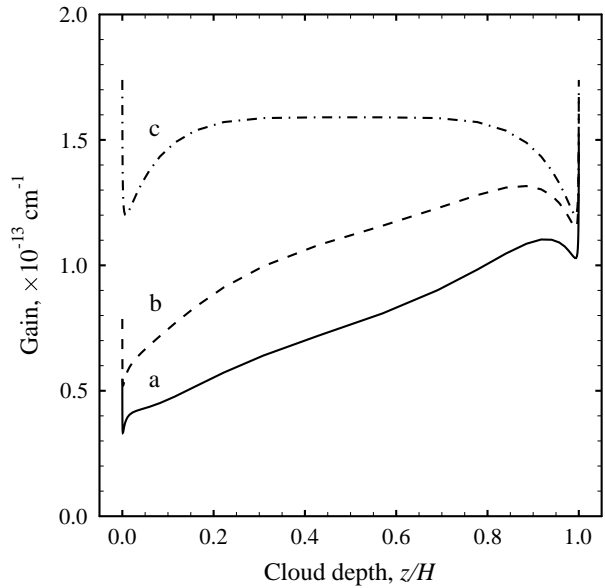}
\caption{The gain of the 658-GHz maser line as a function of the cloud depth. The results are given for three cases: a. The 411 rotational levels belonging to the ground and four excited vibrational states of the molecule are taken into account (solid line). b. The 271 rotational levels belonging to the ground and the first excited vibrational states of the molecule are taken into account (dashed line). c. The 411 rotational levels belonging to the ground and four excited vibrational states of the molecule are taken into account, and the stellar radiation is not taken into account in the radiative transfer (dashed-dotted line).}
\label{fig_depth}
\end{figure}

The dependence of the gain of the 658-GHz maser line on depth into the cloud is shown in the Fig. \ref{fig_depth}. The adopted parameters are $T_\rmn{g} = 1100$~K, $N_\rmn{H_2} = 5 \times 10^{9}$~cm$^{-3}$, ortho-H$_2$O concentration $N_\rmn{H_2O} = 3 \times 10^5$~cm$^{-3}$ and dust--gas mass ratio $f_d = 0.002$, and the Mg$_2$SiO$_4$ dust material was considered. The results are also presented for calculations with smaller number of ro-vibrational levels of the molecule and for calculations without taking into account the stellar radiation in the radiative transfer equation. The differences between calculated results imply that the inclusion of the high-lying excited vibrational states is important in modelling of the emission from the lines belonging to the first excited vibrational state. Analogous conclusions were made by \citet{Gonzalez-Alfonso2007}, \citet{Maercker2009}, \citet{Decin2010a}, who modelled the H$_2$O line emission from the ground vibrational state of the molecule in the circumstellar envelopes of AGB stars. The vibrational excitation of the molecule takes place through collisions and through absorption of stellar radiation. The gain is substantially lower when the stellar radiation is taken into account; the cloud averaged gain $\overline{\gamma}$ at the parameters in question is about $0.8 \times 10^{-13}$~cm$^{-1}$ and $1.5 \times 10^{-13}$~cm$^{-1}$ for the models with and without stellar radiation, respectively (with a full set of ro-vibrational levels). Analogous results were found for the 22.2-GHz circumstellar maser by \citet{Nesterenok2013b}.

\begin{figure}
\includegraphics[width=84mm]{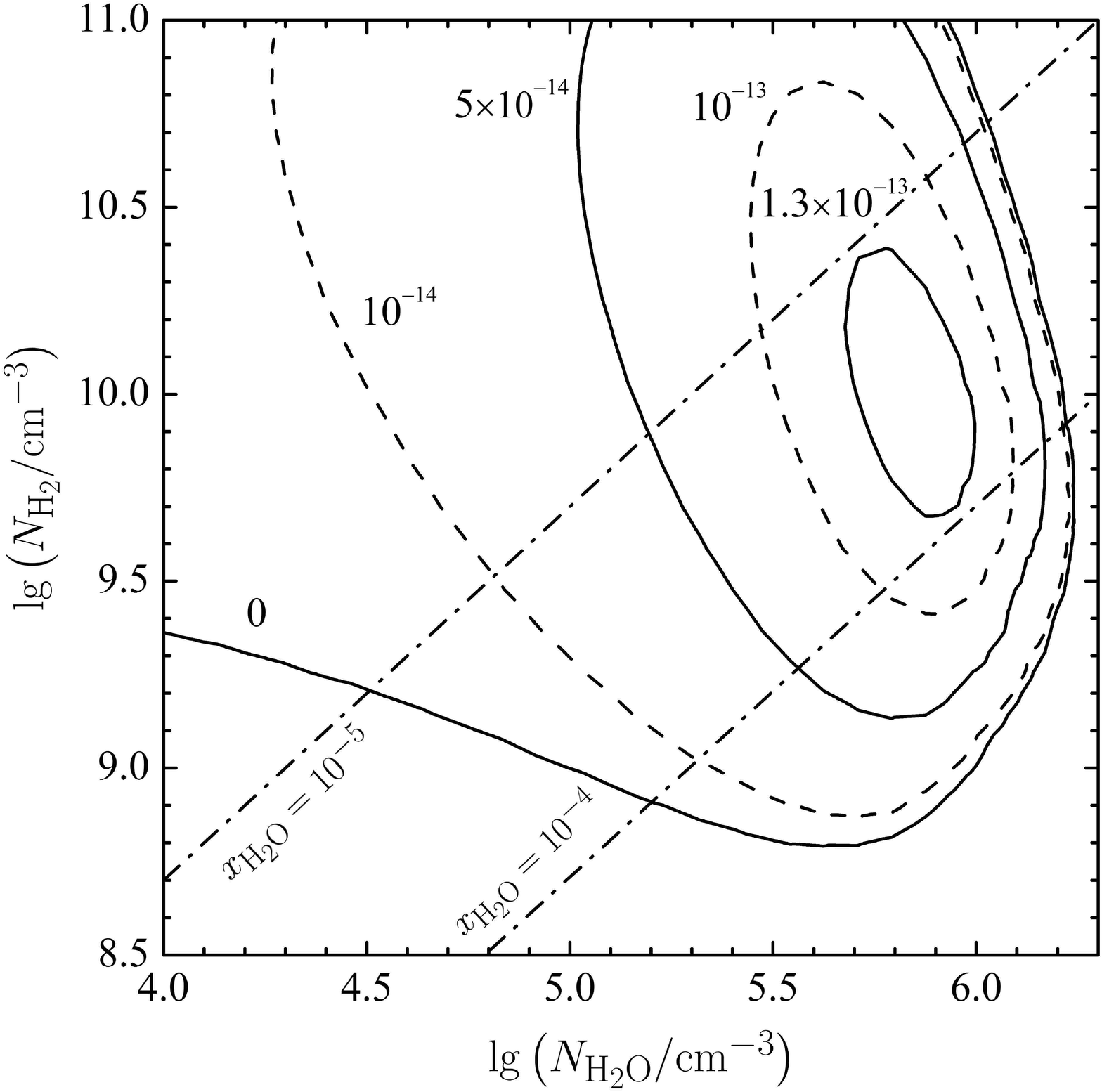}
\caption{The gain of the 658-GHz maser line as a function of the number density of hydrogen molecules $N_\rmn{H_2}$ and the number density of ortho-H$_2$O molecules $N_\rmn{H_2O}$. The contours of equal gain are shown; the gain in cm$^{-1}$ is indicated near each curve. The inclined dash-dot lines correspond to the points for which the relative ortho-H$_2$O abundance $x_\rmn{H_2O} = N_\rmn{H_2O}/2N_\rmn{H_2}$ equals $10^{-5}$ and 10$^{-4}$. The adopted parameters are $T_\rmn{g}$ = 1100~K and $f_d = 0.002$.}
\label{fig_nh2_nh2o}
\end{figure}

Fig. \ref{fig_nh2_nh2o} demonstrates the calculated cloud-averaged gain $\overline{\gamma}$ of the 658-GHz maser line as a function of the number densities of hydrogen molecules H$_2$ and ortho-H$_2$O molecules in the cloud. The gain is $\overline{\gamma} \approx 10^{-13}$~cm$^{-1}$ at H$_2$ number densities about $3 \times 10^9$ to $5 \times 10^{10}$~cm$^{-3}$ and ortho-H$_2$O number densities about $3 \times 10^5$ to $10^{6}$~cm$^{-3}$. The  excitation temperature of the maser levels is $T_\rmn{exc} \sim -10^{3}$~K for this range of physical parameters. The length of the amplification region along the line of sight probably does not exceed, or is comparable to, the distance from the cloud to the star $D$. Thus, the optical depth in the maser line can be as high as $\tau \approx 15$ in the model in question.

The maser becomes saturated if the transition rates between the maser levels are overwhelmingly determined by the induced radiative transitions in the maser line. As the degree of maser saturation increases, the maser luminosity per unit gas volume approaches its limiting value and the line profile broadens \citep{Strelnitskii1975}. The observed contiguous, broad emission of the 658-GHz masers may be interpreted as implying that the masers have a high degree of saturation \citep{Menten1995}. The limiting maser luminosity can be found using the expression given in \citet{Neufeld1991}. In our model, the limiting photon luminosities of the maser $\Phi_\rmn{lim}$ lie in the range $10 - 100$~cm$^{-3}$~s$^{-1}$ at the gas densities and water abundances, at which the gain values are high $\overline{\gamma} > 5 \times 10^{-14}$~cm$^{-1}$. The limiting photon luminosity of the maser cloud is estimated $L_\rmn{lim} \sim \Phi_\rmn{lim} H^3 \sim 10^{41}$~s$^{-1}$ in the model in question (the strong maser emission is suggested to emanate from a cloud predominantly in one direction). We can estimate the beaming angle of a cylindrical maser $\Delta \Omega \sim (H/D)^2 \approx 10^{-2}$~sr. The isotropic photon luminosity is $L_\rmn{is} \approx 4\pi L_\rmn{lim}/\Delta \Omega \sim 10^{44}$~s$^{-1}$ in the model in question. The calculated values of the photon luminosity can roughly reproduce the observed intensities of the maser radiation.

\begin{figure}
\includegraphics[width=84mm]{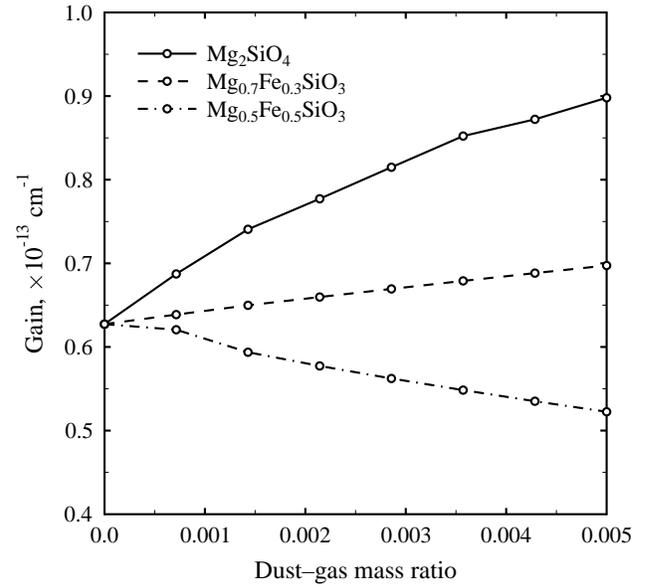}
\caption{The gain of the 658-GHz maser line as a function of the dust--gas mass ratio. The adopted parameters are $T_\rmn{g} = 1100$~K, $N_\rmn{H_2} = 5 \times 10^9$~cm$^{-3}$ and ortho-H$_2$O concentration $N_\rmn{H_2O} = 3 \times 10^5$~cm$^{-3}$.}
\label{fig_dg_ratio}
\end{figure}

Fig. \ref{fig_dg_ratio} shows the dependence of the gain of the maser line $\overline{\gamma}$ on dust--gas mass ratio. The results are presented for three dust materials. The radiative equilibrium temperatures for Mg$_2$SiO$_4$, Mg$_{0.7}$Fe$_{0.3}$SiO$_3$ and Mg$_{0.5}$Fe$_{0.5}$SiO$_3$ dust grains are 340~K, 440~K and 660~K, respectively. The gain of the maser line increases with increasing dust abundance in the case of Mg$_2$SiO$_4$ and Mg$_{0.7}$Fe$_{0.3}$SiO$_3$ dust material. For the Mg$_{0.5}$Fe$_{0.5}$SiO$_3$ dust material the opposite is true -- the gain decreases with increasing dust--gas mass ratio. The radiation of warm dust suppresses the population inversion in the maser line. Analogous results were reported by \citet{Yates1997} for collisionally pumped H$_2$O masers from the ground vibrational state (22.2-, 325-GHz and other lines). The masers that are formed by collisional mechanism are quenched by the strong infrared continuum. This fact explains also the substantial decrease of the gain of the 658-GHz maser when the stellar radiation is taken into account in radiative transfer (see Fig. \ref{fig_depth}).

\subsection{Gain of the other vibrationally excited masers}
The population inversion for other lines  of H$_2$O molecule belonging to the (010) vibrational excited state were considered. The gain of the ortho-H$_2$O transition $4_{14} \to 3_{21}$ at a frequency of 67.8~GHz was found to be $\overline{\gamma} > 10^{-14}$~cm$^{-1}$ at ortho-H$_2$O concentration $N_\rmn{H_2O} > 3 \times 10^5$~cm$^{-3}$ and $T_\rmn{g} = 1100$~K. The maser emission in this line was predicted earlier by \citet{Deguchi1977}. The gain of the ortho-H$_2$O transitions $5_{50} \to 6_{43}$ at a frequency of 233~GHz and $6_{61} - 7_{52}$ at a frequency of 294~GHz was found to be $\overline{\gamma} \ga 10^{-14}$~cm$^{-1}$ at $N_\rmn{H_2O} > 5 \times 10^5$~cm$^{-3}$. There is no population inversion for the ortho-H$_2$O transition $5_{23} \to 6_{16}$ at a frequency of 336~GHz in our model. Note, \citet{Menten2006} observed the 336-GHz line towards VY CMa and concluded, based on the simple considerations, that the line has a thermal excitation. The calculations showed that the population inversion for the transition $1_{10}-1_{01}$ at a frequency of 794~GHz within the (020) vibrational state of the molecule can be high. The gain of the transition is found to be $\overline{\gamma} \ga 10^{-14}$~cm$^{-1}$ at the parameters $N_\rmn{H_2} > 5\times 10^{9}$~cm$^{-3}$, $N_\rmn{H_2O} > 5\times 10^5$~cm$^{-3}$ and $T_\rmn{g} = 1100$~K. This line is a twin of the 658-GHz maser line but is located within the higher excited vibrational state. The calculations did not predict any strong vibrationally excited (sub)millimetre maser for the para-H$_2$O molecule. The gain of the para-H$_2$O vibrationally excited line $4_{40} \to 5_{33}$ at a frequency of 96~GHz was found to be low $\overline{\gamma} \la 10^{-14}$~cm$^{-1}$.

\begin{figure}
\includegraphics[width=84mm]{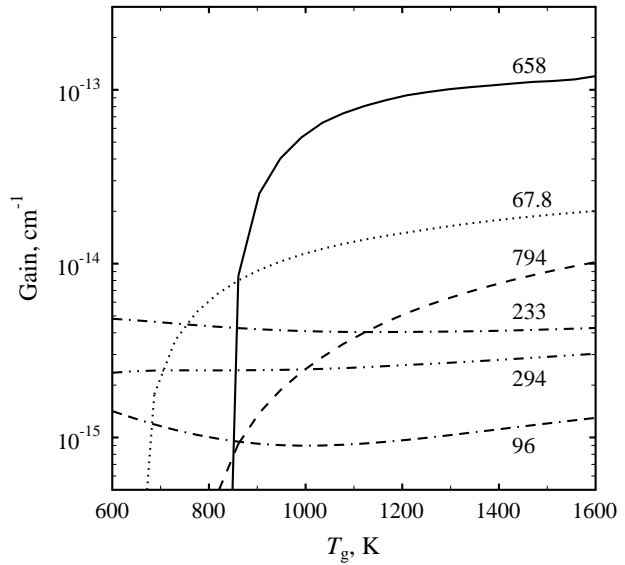}
\caption{The gain of the maser lines as a function of the gas temperature. The line frequency in GHz is indicated near each curve. The adopted parameters are $f_d = 0.002$, $N_\rmn{H_2} = 5 \times 10^9$~cm$^{-3}$, ortho-H$_2$O concentration $N_\rmn{oH_2O} = 3 \times 10^5$~cm$^{-3}$ and para-H$_2$O concentration $N_\rmn{pH_2O} = 10^5$~cm$^{-3}$.}
\label{fig_gas_temp}
\end{figure}

Fig. \ref{fig_gas_temp} presents the dependence of the gain of the vibrationally excited maser lines on gas temperature. The gain depends strongly on gas temperature for the 67.8-GHz, 658-GHz and 794-GHz masers, while for other transitions the gain is almost constant.

According to our calculations, there is weak dependence of the gain of the maser lines on the gas velocity gradient. The variation of the velocity gradient from 0 to 9~km~s$^{-1}$~AU$^{-1}$ leads to about $5-10$ per cent change in the gain of the maser lines in question.  The sensitivity of the calculated gain values to a change in the cloud height was investigated: the change in the cloud height by 10 per cent leads to a 1--5 per cent change in the calculated gain values.

\section{Discussion}

The majority of the H$_2$O masers are collisionally pumped \citep{Yates1997}. The collisional pumping cycle consists of collisional excitation of high-lying ro-vibrational levels of the molecule followed by radiative deexcitation of the levels (the emission of 'sink' photons). The 'sink' photons are absorbed by dust or escape from the resonance region to complete the pumping cycle. The collisional maser pumping is most efficient under conditions when the dust temperature is much lower than the gas temperature \citep{Bolgova1977, Chandra1984, Yates1997}. According to our calculations, the gain of the 658-GHz maser line rises significantly with increasing abundance of cold dust and with increasing gas temperature (see Figs \ref{fig_dg_ratio} and \ref{fig_gas_temp}). These facts indicate that the 658-GHz maser line is collisionally pumped. There is not strong dependence of the gain on the gas temperature for the 96-GHz, 233-GHz and 294-GHz maser lines. Moreover, the gain of these maser lines is substantially lower when the stellar radiation is not taken into account in radiative transfer equation. We suggest that these masers have an important radiative pumping component.

According to the present calculations, gas densities about $10^9 - 10^{10}$~cm$^{-3}$ and relative H$_2$O abundances $10^{-5}-10^{-4}$ are necessary for the effective pumping of the 658-GHz maser. Such values of the water abundance were considered in the models of 22.2-GHz circumstellar masers \citep{Humphreys2001, Nesterenok2013}. The gas density in the stellar wind is about $N_\rmn{H_2} \sim 10^8$~cm$^{-3}$ at distances $D \sim 5$ stellar radii from the star in a model of the circumstellar envelope of oxygen-rich Mira variable with mass loss rate $\dot{M} \approx 7 \times 10^{-6}$~M$_{\odot}$~year$^{-1}$ \citep{Jeong2003}. The gas density and water abundance in the maser clouds can differ substantially from the values calculated in the framework of the isotropic circumstellar wind \citep{Richards2012, Nesterenok2013}.

The gas temperature in the maser cloud must be sufficiently high for the effective collisional excitation of ro-vibrational levels of the H$_2$O molecule. At the densities $10^9 - 10^{11}$~cm$^{-3}$, the efficiency of the energy exchange between gas and radiation field is at maximum which results in a close coupling of the gas to the condition of radiative equilibrium \citep{Schirrmacher2003}. The 658-GHz maser emission in the circumstellar envelope of supergiant VY CMa is found to arise from an extended region at distances up to 40 stellar radii from the star \citep{Richards2014}. The radiative equilibrium temperatures are too low for the pumping of 658-GHz masers at such distances. Thus, the local heating mechanisms must play a significant role in the gas energetic equilibrium in the maser clouds. These gas heating mechanisms can be H$_2$-formation \citep{Schirrmacher2003}, dissipation of turbulent motions of the gas \citep{Strelnitski2002}, dissipation of magneto-hydrodynamic and acoustic waves \citep{Pijpers1989}. The gas temperature in the post-shock layers, where the hydrogen molecule formation takes place, can be higher by 300 -- 400~K than the radiative equilibrium temperature \citep{Schirrmacher2003}. Note that some maser emission in VY CMa originates from elongated features characteristic of post-shock gas \citep{Richards2014}. 

Only those dust particles whose radiative equilibrium temperature is lower than the condensation temperature (which is about 1000 –- 1100~K for silicates) can survive \citep{Gail2010}. The strong candidates for dust material in the inner layers of circumstellar envelopes of oxygen-rich AGB stars are silicates with poor Fe abundance
\citep{Bladh2012}. Fe-free silicate material is more efficient at emitting the radiation than absorbing it. The relatively cold Fe-free silicate dust makes the collisional pumping of the masers efficient.

The time variation of H$_2$O circumstellar masers in 22.2-GHz line is generally attributed to a change of the gas density and the velocity field of a circumstellar envelope \citep{Deguchi1977, Rudnitskii2005, Shintani2008}. We suggest that the variation of the infrared stellar luminosity can affect significantly the maser luminosities because of a strong coupling of the gas to the condition of radiative equilibrium. The lower the stellar luminosity, the lower the gas temperature and the less effective the collisional pumping of the masers. Moreover, the strong infrared radiation field affects the pumping mechanism and lowers the gain of the collisionally pumped masers (for radiatively pumped masers the effect has an opposite sign). It is believed that the inner radius of the H$_2$O maser emission zone in a circumstellar envelope is determined by the distance from the star where the gas density in the maser cloud falls below the collisional quenching density for the maser \citep{Cooke1985, Richards2012}. We suggest that the stellar radiation is another factor, that can affect significantly the maser pumping mechanism and quench the collisionally pumped masers in the inner layers of circumstellar envelopes of late-type stars. 

\section{Conclusions}
We investigate the physical conditions necessary for the generation of the vibrationally excited H$_2$O maser at 658~GHz in the inner layers of the circumstellar envelopes of AGB stars. It is shown that the high-lying excited vibrational states and the stellar radiation have to be included in the calculations of the level populations belonging to the first excited vibrational state. There is strong dependence of the gain of the 658-GHz maser line on gas temperature and on dust composition and abundance. The maser is found to be collisionally pumped. 

The gain of other maser transitions belonging to excited vibrational states were calculated. The ortho-H$_2$O masers at 67.8~GHz and 794~GHz were found to be collisionally pumped. The ortho-H$_2$O masers at 233~GHz and 294~GHz and the para-H$_2$O maser at 96~GHz were found to have an important radiative pumping component. We suggest that the maser transition at 67.8~GHz might prove to be detectable from the ground, despite the fact that this maser was searched for earlier without success.

It will be of particular interest to monitor observations of the stellar 658-GHz maser and other vibrationally excited maser lines and to determine of the time variability properties of the H$_2$O masers.

\section*{Acknowledgements}
This work was supported by the Russian Foundation for Basic Research (project no. 14-02-31302), the Program of the President of Russia for Support of Leading Scientific Schools (project no. NSh-294.2014.2).  

\bibliography{references_the_model_of_maser_658_GHz}

\end{document}